\documentclass[12pt]{article}
\usepackage[T1]{fontenc}
\usepackage[dvipsnames]{xcolor}
\usepackage{amsmath,amssymb}
\usepackage{graphicx}
\usepackage{placeins}
\usepackage{float}
\usepackage{algorithm}
\usepackage{algpseudocode}
\usepackage{mathtools}
\usepackage{booktabs}
\usepackage[round,authoryear]{natbib}
\let\textcite\citet
\let\parencite\citep
\let\autocite\citep
\usepackage[hidelinks]{hyperref}
\usepackage{microtype}
\title{Forecasting Extreme Day and Night Heat in Paris: A Proof of Concept}
\author{Richard Berk\\University of Pennsylvania (Emeritus)\\
\texttt{berk@sas.upenn.edu}}
\date{}
\begin{document}
\maketitle
\begin{abstract}
As a form of ``small AI'', quantile machine learning is used to forecast diurnal and nocturnal $Q(.90)$ air temperatures for Paris, France from late spring through the summer months of 2021. The data are provided by the Paris-Montsouris weather station. Rather than trying to directly anticipate the onset and cessation of reported heat waves, $Q(.90)$ values are estimated. The 90th percentile is chosen so that exceedances represent relatively rare and extreme conditions. Predictors include eight routinely available indicators of weather conditions, lagged by 14 days. Using holdout data, the temperature forecasts are produced two weeks in advance. Adaptive conformal prediction regions are computed that, under exchangeability, provide provably valid finite-sample coverage of forecasting uncertainty. For both diurnal and nocturnal temperatures, forecasting accuracy in the holdout data is promising, and sound measures of uncertainty are coupled with a novel decision-making framework. Benefits for policy and practice follow.
\end{abstract}

\section{Introduction}

Anthropogenic global warming has long been recognized \citep{Schneider1989}. More recent concerns center on associated increases in the frequency and intensity of unusually hot weather at local scales. \citep[chap.~13]{Tziperman2022}. These changes produce inordinate impacts on ecosystems \citep{Stillman2019,Breshears2021} and public health \citep{Ballester2023,Cvijanovic2023}.

Accurate forecasts of rare, high temperatures offer significant benefits for subject-matter understanding. Policy preparedness can benefit as well \citep{Xu2014,Pascal2021}. There are impressive data analyses and simulations that help, but implementation costs are high and performance often degrades at the smaller spatial scales where such forecasts are most needed. Valid uncertainty estimates are commonly lacking.  All three deficiencies can undermine scientific understanding and public policy. In this paper, computational burdens, appropriate spatial scales and valid uncertainty estimates are constructively addressed.

Forecasting extreme heat is undertaken by applying quantile machine learning along with adaptive conformal prediction regions to weather station data \citep{Romano2019}. $Q(.90)$ diurnal and nocturnal temperatures are forecast because such temperatures are by construction extreme and rare. The statistical approach can be seen as a complement to the ``industrial strength'' methods that seem to dominate the literature. Implications directly follow for early heat warning systems at instructive local scales.

Section 2 briefly provides some statistical background on past heat forecasting studies to motivate the later data analysis and forecasts. Section 3 describes the data and forecasting methods. Section 4 presents some descriptive information, summaries of the training data results, and $Q(.90)$ temperature forecasts from holdout data informed by adaptive conformal prediction regions. Section 5 is a discussion of the results, their implications for policy and practice, and for proposed future work. Conclusions are drawn in Section 6.

\section{Statistical motivation}

Climate science commonly provides a foundation for forecasts of rare, high temperatures \citep{Petoukhov2013, Mann2018,McKinnon2022, Li2024}. In an instructive review, \citeauthor{Domeisen2023}~\citeyearpar{Domeisen2023} write,
\begin{quote}
Understanding of the processes influencing heatwave development and characteristics enables improved representation in models, thereby enhancing long-range prediction capabilities. These processes include those from the atmosphere as well as the land or ocean surface encompassing drivers (large-scale local and remote processes communicated to the heatwave location as changes in temperature, humidity and circulation) and feedbacks (a combination of regional-scale processes of mutual influence on a subcontinental scale).
\end{quote}

When a physics-informed model is sufficiently complete and correct, accurate forecasts can be a useful byproduct. But even very good subject-matter models may lack capabilities that matter for certain applications. For example, the widely used Community Earth System Model (CESM) seems ill-equipped to address rare and extreme heat, especially at the small scales often required.\footnote
{
The standard grid size for the atmosphere and land components of the CESM is nominally \(1^{\circ}\) of latitude by \(1^{\circ}\) longitude or for mid latitude locations, roughly 100~km by 100~km (i.e., 10,000 squared km). In climate science, this is in the meso-scale range. For the weather station data used here, there is no explicit grid, but a reasonable grid for many locations would be about 10~km by 10~km, which is in the local scale range \citep{Oke1987}.
}
Recent research suggests that deep learning might improve downscaling \citep{Wang2021}, although a new and complicated statistical overlay would be added.

The CESM also has difficulty properly accounting for uncertainty. \citeauthor{Gettleman2016}~\citeyearpar{Gettleman2016} summarize the issues: ``Uncertainty in climate models has several components. They are related to the model itself, to the initial conditions of the model \ldots\ and to the inputs that affect the model \ldots\ All three must be addressed for the model to be useful.'' Were this accomplished, along with successful downscaling, the CESM might be closer to producing a credible \emph{distribution} of outcomes that might capture rare climate events in its tails.

Finally, the CESM depends on costly high-performance computing. It has a large, parallel, Fortran codebase configured for supercomputers or large clusters. Simulations of century-scale, coupled climate processes can require thousands of cores and massive memory. Even small subsets of the code (e.g., atmosphere only) typically require clusters or cloud access \citep{NCAR2025}.

Algorithmic methods are a complementary approach that can provide useful forecasts at smaller spatial scales combined with valid estimates of uncertainty and substantial computational savings. However, an algorithm is not a model \citep{Breiman2001}. As \citeauthor{Kearns2019} \citeyearpar{Kearns2019} emphasize, ``At its most fundamental level, an algorithm is nothing more than a very precisely specified series of instructions for performing some concrete task.'' Algorithms are evaluated by how well they accomplish that concrete task, not by how well they represent known physics or other sciences, nor by their explanatory power.''  

Yet, recent work shows some of the promise in algorithmic approaches used to forecast salient climate events. For example, \citeauthor{Bodnar2025}
\citeyearpar{Bodnar2025} build a large-scale machine learning procedure, trained on ``earth system'' data. The algorithm can be ``fine tuned'' for particular forecasting applications at appropriate temporal and spatial scales. Hurricane tracking is one instance for which the forecasting results are impressive. However, the requisite training is a massive computational undertaking, and even the fine tuning requires substantial data processing power and human capital. In addition, forecasting uncertainty has yet to be addressed; the researchers seem committed to using forecasting ensembles whose formal statistical properties are unspecified and perhaps problematic \citep{Fu2025}.

A commitment to algorithmic forecasting does not preclude procedures that can cross the algorithm--model barrier. Particular features of climate science can be incorporated \citep{Hao2022}. For example, constraints extracted from thermodynamics can be imposed on a neural network.  These enhancements are intended primarily to improve algorithmic performance. This is an important feature of the analyses below.\footnote
{
There are also applications in which climate simulations, enhanced by statistical procedures, are used to further explanation and understanding \citep{Fischer2023}. Statistical enhancements of numerical weather prediction can be seen in this manner \citep{Price2024}.
}

\subsection{Particular challenges for algorithmic high temperature forecasts}

Difficulties in the training and use of machine learning algorithms are widely
discussed by computer scientists \citep{Goodfellow2017} and statisticians \citep{Hastie2009}. There are two problems for algorithmic temperature forecasts that help inform the data collection and analyses to follow.

First is a tendency to focus on heat waves as binary events. The mechanisms governing extreme heat are increasingly understood \citep[chap.~13]{Tziperman2022}, but forecasting the presence or absence of heat waves, rather than high temperatures, can be a distraction \citep{Smith2013}. \citeauthor{Perkins2013} \citeyearpar{Perkins2013} caution, ``\ldots\ definitions and measurements of heat waves are ambiguous and inconsistent, generally being endemic to only the group affected, or the respective study reporting the analysis.'' Moreover, heat wave definitions can be media-driven \citep{Hulme2008,Hopke2020}. Noteworthy heat is newsworthy heat. In short, it can be risky to treat heat waves as discrete physical events when the reality is far more nuanced.

Second, the role of excessive nocturnal heat commonly is overlooked. Yet, high nocturnal temperatures can substantially threaten local ecosystems and public health. Critical recovery time from excessive daytime temperatures can be sacrificed \citep{Walther2002,Anderson2009,He2022}. Nocturnal temperatures are easy to neglect because they are almost never the highest daily temperatures. In addition, they are shaped by somewhat different mechanisms from diurnal temperatures, making them especially challenging to forecast.

\section{Data and methods}

Weather station data can be seen, in part, as a response to spatial scales that are too coarse. The data used were provided by the Paris--Montsouris weather station. Observations from 2020 constitute the training data. Holdout observations from 2019 are used for calibration, while holdout observations from 2021 are used for forecasting \footnote
{
2020 is bracketed by 2019 and 2021 so that 2019 and 2020 are as close as possible in time to the training data.
}
A temporal index \(t = 1,2,3,\ldots, T\) denotes each of 214 days from March1 to September~30 when unusually warm temperatures can occur.\footnote
{
Data from March were included primarily to obtain the values of the lagged predictors for each of the corresponding early days in April.
}
Days are a common temporal unit for studies of unusually high temperatures.

Paris is the study site. Any of several other locales could have been selected and will be in future work. Paris currently is perhaps Europe's urban, high temperature ground zero \citep{Porter2025}, arguably with Europe's most heat-vulnerable urban population \citep{Masselot2023}.

The two response variables are centigrade air temperatures at 2~PM and 2~AM solar time. Solar time provides a useful and consistent time stamp while avoiding local conventions such as daylight saving time. The 2~PM and 2~AM temperatures do not necessarily represent the most extreme diurnal or nocturnal heat day after day but serve as reasonable proxies. Fixed measurement times also avoid heat effects that vary substantially across the day. For example, a peak temperature at noon will have different effects on outdoor workers than a peak temperature at 3~PM because of breaks taken for lunch in the middle of the day. Summary statistics such as the daily mean temperature are sometimes used, but risk averaging out extreme heat in the right tails of temperature distributions.

Measured temperatures rather than Steadman heat index values are favored for the response variables because of well-known problems with the Steadman heat index at temperatures below \(80^{\circ}\)F \citep{Steadman1979,Rothfusz1990}. During the summer months, such temperatures are common in Paris after dark. There is a good chance of nonsense results.

Predictors are limited to information readily available in weather station data. They are lagged here to identify the direction of any causal ordering and to provide stakeholders a warning in advance of impending extreme heat. The lag of 14 days is imposed, consistent with earlier research on the 2021 Pacific Northwest (i.e., North American) heat wave \citep{Li2024}. 

The eight lagged predictors include: (1) wind direction in degrees from true north, (2) wind speed in meters per second, (3) air temperature in degrees Celsius, (4) atmospheric pressure in hectopascals (hPa), (5) visibility in meters, (6) dew point in degrees Celsius, (7) relative humidity in percent units, and (8) a counter for the day. The counter is included to capture temporal trends. On average, early August will be warmer than early June, although the increases can be nonlinear over time.
At least some of the predictors are likely to be related in complicated ways to well-known precursors of certain excessive heat regimes. For example, dry soil, the absence of clouds, and elevated barometric pressure in the mid-troposphere sometimes contribute to high-order interaction effects with routine seasonal warming \citep[chap.~13]{Tziperman2022}.

Wind direction requires a transformation before use as a predictor. Wind direction is a circular variable measured in degrees, with $0^{\circ}$ and $360^{\circ}$ representing the same physical direction. Treating wind direction as a linear predictor can therefore induce artificial discontinuities near the wrap-around point. To address this, one can transform wind direction for $\theta_t \in [0,360)$ using its sine and cosine components,
\[
\mathrm{wd}_{\sin,t} = \sin\!\left(\frac{2\pi \theta_t}{360}\right),
\qquad
\mathrm{wd}_{\cos,t} = \cos\!\left(\frac{2\pi \theta_t}{360}\right).
\]
This transformation embeds wind direction on the unit circle, ensuring that directions close in angle (e.g., $359^{\circ}$ and $1^{\circ})$ are also close in predictor space. The pair $(\mathrm{wd}_{\sin,t}, \mathrm{wd}_{\cos,t})$) preserves directional information without imposing an arbitrary origin and allows standard regression and machine-learning methods to fit appropriate directional effects. When the original wind direction variable is replaced by the two trigonometric functions, there are 9 lagged predictors rather than 8.

\subsection{Temporal Dependence}

The 2~AM and 2~PM response variables combined with the 9 predictors constitute a multiple time series. Because of the data's longitudinal structure, temporal dependence can create two important complications. First, holdout data obtained by random sampling will scramble time series dependence \citep[sec.~5.8]{Hyndman2021}. As an alternative, calibration data for adaptive conformal prediction regions are drawn from the Paris--Montsouris weather station from March~1st through September~30th, 2019. ``Honest'' forecasts are obtained from other holdout data drawn from the Paris--Montsouris weather station from March~1 to September~30, 2021. The same physical processes should apply to the corresponding months in 2019, 2020, and 2021, although there can be significant random variation in the realized data. These issues are empirically addressed later as they arise.\footnote
{
Some of the issues can be subtle. Important predictors might be concentrated in very different regions of the predictor space in different seasons. With strong nonlinear relationships \citep[chap.~3]{Stull2017}, predictor values might fall at relatively flat parts of the response function in winter and at relatively steep parts of the response function in the summer (or vice versa). Yet the response function is the same. As an empirical matter, this might look like a change in the response function itself. Because forecasting, not explanation, is the intent, such complications can be postponed.
}

Second, for the 2~PM temperatures, the multiple time series observations are analyzed with quantile gradient boosting \citep{Friedman2002} using a .90 quantile ($Q(.90)$) estimation target to focus on extreme and rare high temperatures \citep{Velthoen2023}. Using quantiles also sidesteps the need to rely on reported heat wave definitions. However, temporal dependence can undermine calibration data exchangeability required for conformal prediction regions \citep{Angelopoulos2025}. For the 2~AM temperatures, the multiple time series observations are analyzed in a somewhat more complicated manner arising from nocturnal boundary layer processes, and uncertainty estimation can be similarly compromised. Valid estimates of 2~PM and 2~AM forecasting uncertainty motivate additional steps to remove temporal dependence from the calibration data \parencite{Chernozhukov2018}. The approach used is best discussed when the forecasting results are addressed.

\subsection{Forecasting Evaluation}

The combination of quantile gradient boosting and conformal inference can work well in concert despite substantial temporal dependence \emph{in training data}. However, many common forecasting assessment tools are no longer valid \citep{Koenker1999}. For example, the Brier Score is not designed for forecast quantiles. A quantile scoring rule is needed. Related features of quantiles invalidate the false alarm ratio, the probability of detection, directional accuracy, the mean squared or mean absolute error, and the AUC-ROC. The forecasts reported below build on the pinball loss function for quantile gradient boosting and on conformal prediction coverage and precision. More detail on both is provided later.

\section{2020 Training Data Results}

\subsection{Response variable descriptive statistics}

\begin{figure}[t]
 \includegraphics[width=0.8\textwidth]{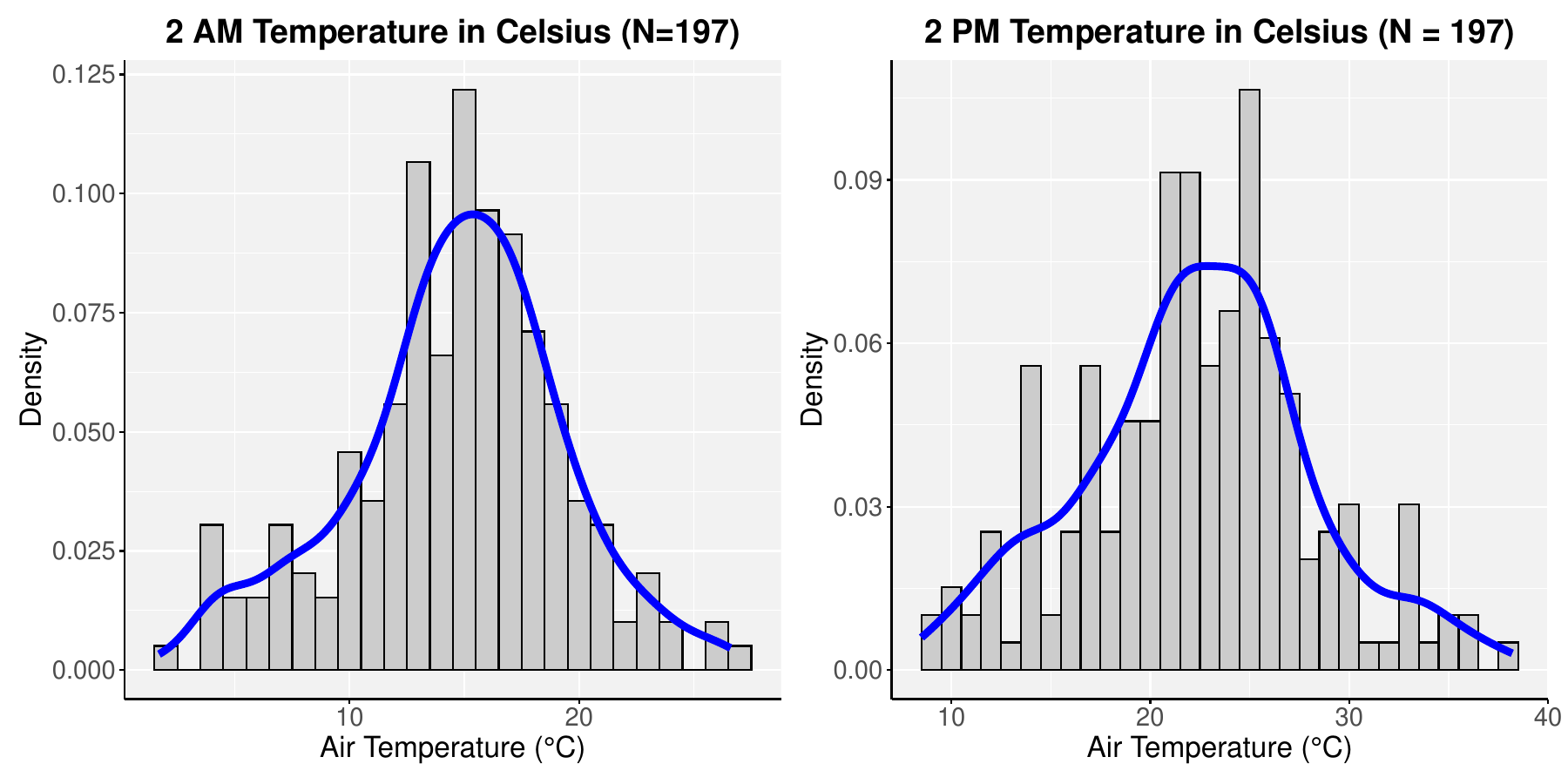}
  \caption{Histograms of the Paris daily 2~AM air temperatures in the left panel and 2~PM air temperatures in the right panel, both in Celsius, for April 1 through September 30 in 2020. The solid blue line in both panels is an overlaid density smoother serving as a visual aid. ($N = 183$ days)}
  \label{fig:hists}
\end{figure}

Figure~\ref{fig:hists} displays on the left a histogram of 2~AM Celsius air temperatures with a density smoother overlaid. The right histogram provides the same information for the 2~PM Celsius temperatures. Both histograms are approximately symmetric with about the same spread and and do not exhibit the long right tail characteristic of the generalized extreme value distribution emphasized by some researchers.  As expected, the 2~PM temperatures tend to show higher values. The 2~PM $Q(0.90)$ value of approximately $30^{\circ}$C. The 2~AM $Q(0.90)$ value is approximately $20^{\circ}$C.

The 0.90 quantile provides a statistical definition of \textit{rare}. It represents a compromise between a focus on atypical temperatures and the need for important regions in the predictor space to contain sufficient data. For both distributions, their right tails include several relatively high temperatures. None appear as obvious outliers. These values are potential forecasting targets, but they arise from marginal distributions; conditional distributions are required for forecasting.

\subsection{Fitting the 2~PM temperatures}

Fitting the $Q(0.90)$, 2~PM temperatures with quantile gradient boosting implies an asymmetric loss function that incentivizes the boosting algorithm to weight underestimates far more heavily than overestimates. This is a desirable feature not a bug. For the quantile loss function used, $\tau = 0.90$. The pinball loss is: 
\begin{equation}
L_\tau(y,\hat y) =
\begin{cases}
\tau\,(y - \hat y), & \text{if } y \ge \hat y,\\[4pt]
(1-\tau)\,(\hat y - y), & \text{if } y < \hat y.
\end{cases}
\label{eq:quantileloss}
\end{equation}
As a result of the specified value of $\tau$, underestimates are 9 times more costly in the loss than overestimates.

\begin{figure}[t]
  \centering
  \includegraphics[width=0.9\textwidth]{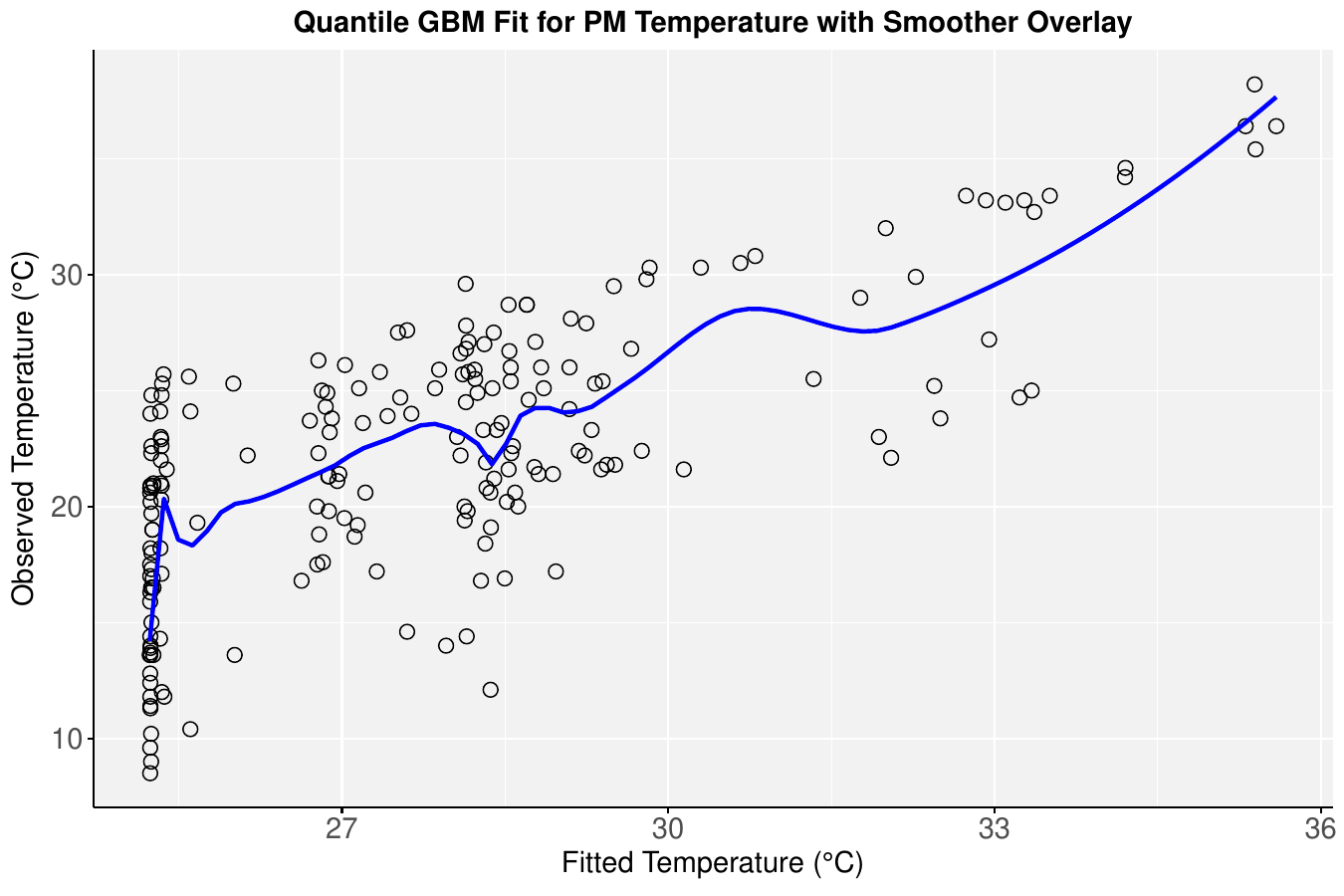}
  \caption{Fit quality is displayed for quantile gradient boosting applied to the 2020 training data daily 2~PM temperatures. The vertical axis represents the observed temperatures in Celsius. The horizontal axis represents the $Q(0.90)$ fitted temperatures in Celsius. The open circles are the observations, and the solid blue line is a loess smooth provided as a visual aid. ($N = 183$ because of lagging.)}
 \label{fig:Fit}
\end{figure}

Figure~\ref{fig:Fit} is a plot of the 2020 observed 2~PM Celsius temperatures against the 2020 2~PM fitted $Q(0.90)$ Celsius temperatures. The fitted values are a produced by the trained quantile boosting algorithm with all nine predictors measured two weeks earlier; the afternoon temperatures are anticipated 14 days in advance.

The \texttt{gbm} procedure in \textsf{R} was used \citep{Ridgeway2024}. At large quantiles, sparse data can be anticipated. Consequently, the shrinkage value was specified as 0.0001 to encourage gradual improvements over iterations. Interaction depth was set to 6 to capture rare, complex interactions. The minimum node size was set to 5 to encourage ``deep'' trees.

The relationship in Figure~\ref{fig:Fit} is approximately linear and positive with local positive and negative departures.The overall trend is not surprising, and serves as a sanity check for the fitting approach used; as fitted temperature increase, their observed temperatures  increase as well. The local variation suggests that beyond a linear trend, there are some localized processes pushing the fitted values up or down. The curious vertical cluster for the observed temperatures at the lowest fitted value results from fitting $Q(0.90)$; by design, the fitted values tend to fall above the bulk of the data and miss low temperature variation. This is one important reason why formal measures of fit can be misleading for quantile estimation \citep{Koenker1999}, especially when the quantile is some distance from the median.

A plot of the each variable's influence on the fitted values is dominated by the day counter, which captures slowly evolving seasonal trends, though all lagged predictors contribute as well.  Partial dependence plots show that the conditional relationships between the lagged predictors and the response variables are generally quite nonlinear. Both additional displays of the boosting results \citep{Friedman2001,Friedman2002} are a secondary concern here because, again, an algorithm is not a model. Those results are omitted for brevity.

\subsection{Time Series Display for the 2020 2~PM Training Data Temperatures}

\begin{figure}[h!]
\centering
 \includegraphics[width=0.9\textwidth]{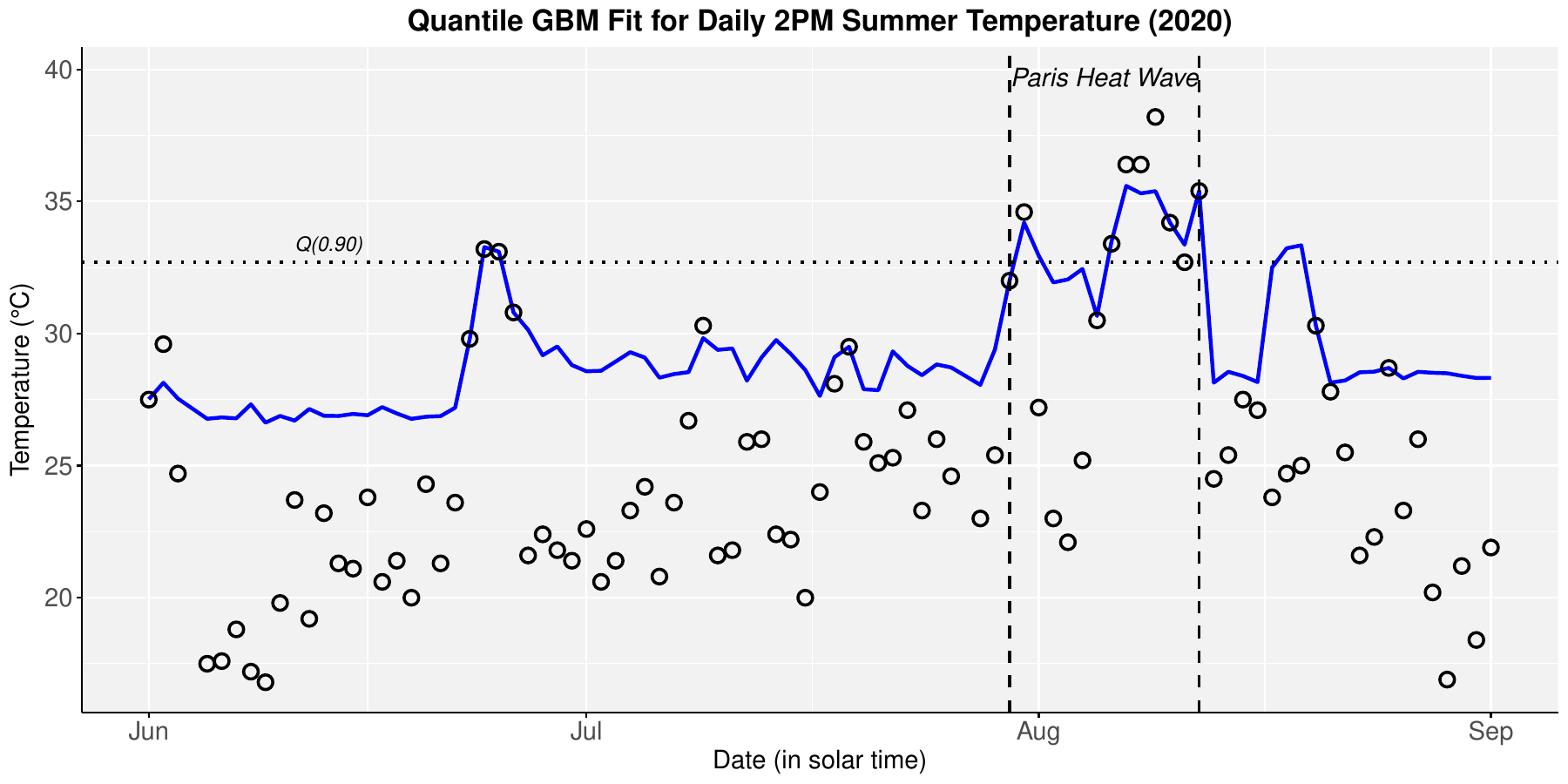}\\
  \caption{A time series plot is shown with the training data 2~PM Celsius temperatures on the vertical axis, date on the horizontal axis, and a loess smooth of the fitted values overlaid to help visualize the temporal path of the fitted values (span = 0.30). The circles are the observations. The horizontal dotted line is placed at the $Q(0.90)$ value of the June through August data. To help avoid clutter, only the summer months are shown. If there is excessive heat, these are the months when it is most likely. The two vertical lines mark a reported heat wave.}
 \label{fig:TS}
\end{figure}

An instructive display can be constructed by plotting the same data responsible for Figure~\ref{fig:Fit} reorganized to highlight trends over time. Figure~\ref{fig:TS} shows the result. By construction, the \texttt{gbm} fitted 2~PM temperatures in Figure~3 tend fall above the observed temperatures because of the quantile loss function with $\tau = 0.90$. There are no strong temporal trends in the fitted values over the subset of months included in the figure, but several peaks fall above the observed temperature 90th percentile.\footnote
{
There are evident temporal trends when the months of April, May and September are included in a similar figure. It is cooler before April and after September.
}

There is one prominent peak in the fitted values that corresponds well to a reported heat wave beginning on July~28th and ending on August~13th. Knowledge of its existence and timing played no role in the analysis but serves as external evidence that the fitted values can capture sustained heat regimes.\footnote
{
The existence of the heat wave was determined through an internet search after the data analysis was completed. The Copernicus Climate Change Service, which is a well respected scientific organization, was the source of the heat wave information.
}
Because the predictors are lagged by 14 days, they lead the reported heat wave by two weeks. However, the correspondence is a function of the \texttt{gbm} fitted values from 2020 training data. True forecasting skill is addressed shortly.

\FloatBarrier

\subsection{Fitting the 2~AM temperatures in the 2020 Training Data} 

Fitting the 2~AM temperatures in the training data requires a modification to the fitting procedure that reflects the physical mechanisms governing nocturnal temperatures.\citep{Oke1987}. During daylight hours, the net rate of energy gain near the earth's surface is dominated by solar heating of that surface, which in turn drives turbulent mixing within the atmospheric boundary layer.\footnote
{
The boundary layer is the part of the troposphere where the effects of surface friction, surface heating and cooling, moisture fluxes, and surface roughness generate turbulent motions on time scales of about an hour.
}
Under these conditions, the surface and the near-surface atmosphere are said to be coupled.

During the evening transition, the near-surface atmosphere remains weakly coupled to the overlying air through residual turbulence. As radiative cooling proceeds, a stable temperature inversion typically forms near the surface, suppressing turbulent exchange and leading to nocturnal decoupling. By the early morning hours (e.g., 2~AM), turbulent mixing is weak or intermittent, and near-surface air temperatures reflect a combination of radiative cooling and the thermal inertia of the surface--atmosphere system, together with the influence of large-scale air-mass conditions from the preceding day.

As a consequence, the systematic evolution of nighttime temperatures depends depends in a systematic way on the prior daytime thermal state, including sustained multi-day anomalies such as extreme heat, which can raise both afternoon and nighttime temperatures over extended periods. This motivates representing the baseline afternoon--nighttime relationship using a nonparametric smoother that captures a slowly varying conditional structure,
\begin{equation}
T_{2am,t} = f(T_{pm,t}) + \eta_t.
\label{eq:am_pm}
\end{equation}
In practice, the function $f(\cdot)$ can be estimated using a
data-adaptive procedure that targets systematic upper-tail nighttime
behavior rather than average conditions. As such, it reflects the influence of
persistent thermal anomalies during warm periods. The deviation term
$\eta_t$ represents higher-frequency nocturnal variability arising from
processes not summarized by afternoon temperature alone, including
night-to-night changes in cloud cover and other transient effects. These
processes can generate the sharp day-to-day peaks and valleys observed around the fitted baseline, while leaving the broader heat-wave--scale structure
intact.\footnote
{
The decomposition in Equation~\eqref{eq:am_pm} is introduced here to motivate the baseline afternoon--nighttime relationship and its sources of variability; no assumptions are made at this stage about the distributional properties of the deviation term $\eta_t$. But temporal dependence can be anticipated. Details are provided in the pseudocode appendix.
}

In this study, the baseline function $f(\cdot)$ was estimated using additive quantile regression smoothing, which extends classical regression smoothing to conditional quantiles \citep{Koenker1994}. Rather than modeling the conditional mean of nighttime temperature given the prior afternoon temperature, this approach targets the upper conditional tail ($\tau = 0.90$), which is more directly relevant for sustained warm nights when there is extreme heat during the day. The resulting smooth captures the slowly varying, thermally driven component of the afternoon--nighttime relationship, while allowing sharper day-to-day deviations to be absorbed by the residual term $\eta_t$.

In summary, the physical mechanisms underlying these temperature processes are well understood and described in standard meteorological texts \citep{Oke1987,Stull2017}. The governing relationships are typically expressed as systems of differential equations involving radiative fluxes, turbulent transport, and thermodynamic state variables. These quantities are not directly observed in routine weather-station data and are therefore not empirically resolvable at the multi-day lead times considered here. Consequently, $f(T_{pm,t})$ can be viewed as a \emph{data-driven approximation to the time-integrated effects of more fundamental physical properties}.

\subsection{Time series display for the 2020 Training Data 2~AM Temperatures}

Training for the 2~AM temperature forecasts proceeded in two steps. First, quantile gradient boosting, fitted earlier, was used to obtain projected 2~PM temperatures at the $\tau = 0.90$ quantile for the two-week forecasting horizon. Second, these projected 2~PM quantile values were used as a predictor of the corresponding 2~AM temperatures 12 hours later.\footnote
{
The observed 2~PM temperature from the preceding day cannot be used because it would not be available in real time when the forecasting is undertaken.
}
The relationship between projected 2~PM temperatures and observed 2~AM temperatures was estimated using a quantile regression smoother, computed with the \texttt{rqss} function in the \texttt{quantreg} \textsf{R} package \citep{Koenker2017}. The smoother targets the $\tau=0.90$ conditional quantile, capturing the slowly varying baseline of the afternoon--nighttime relationship while allowing sharper day-to-day deviations to be absorbed by the residual process."

\begin{figure}[t]
\centering
 \includegraphics[width=0.9\textwidth]{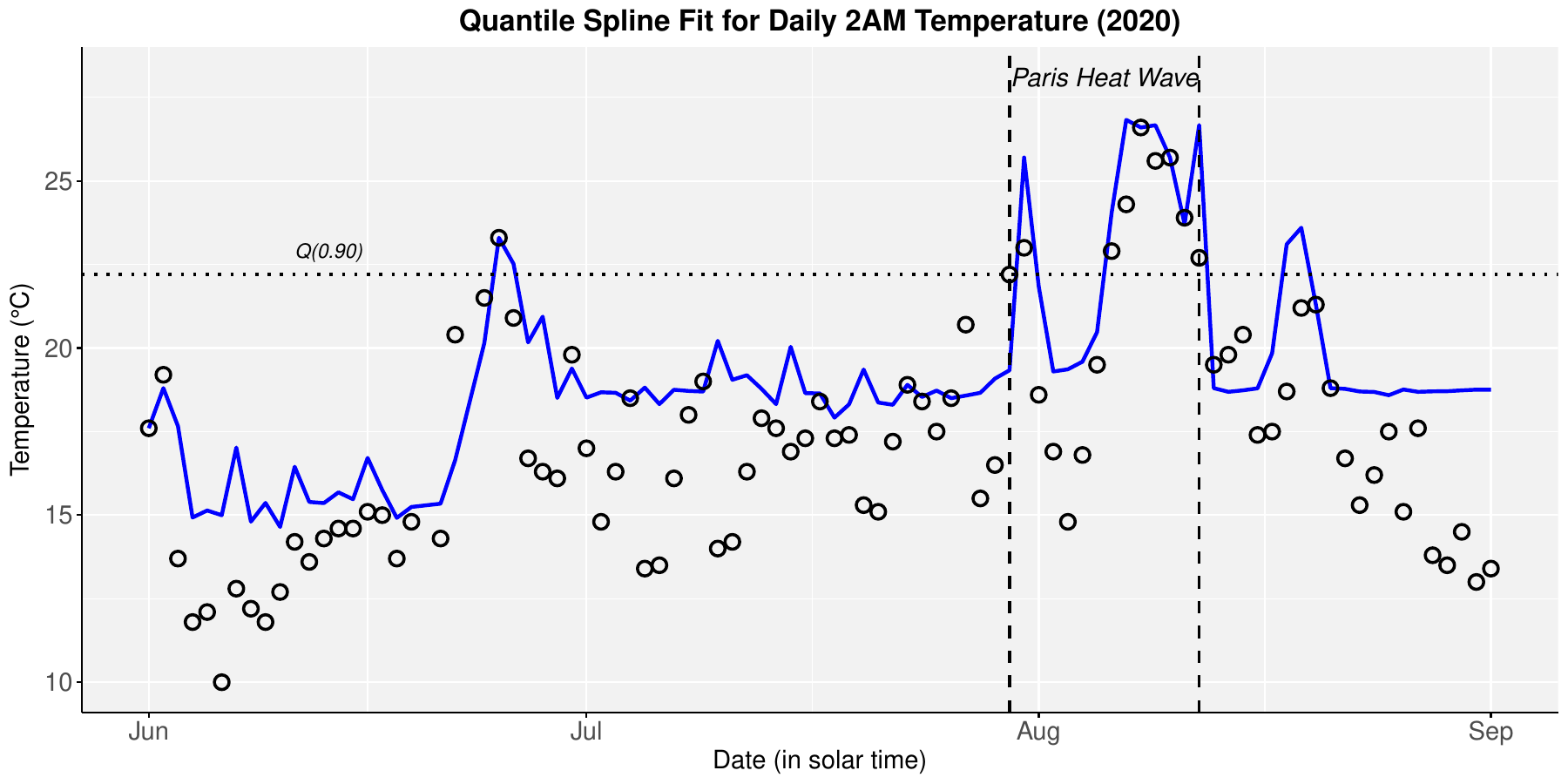}
  \caption{The figure shows 2020 time series of observed 2~AM temperatures (circles) with an overlaid $\tau = 0.90$ quantile regression smoother fit based on projected 2~PM temperatures. The horizontal dotted line marks the empirical $Q(0.90)$ threshold computed from June through August observations. To reduce visual clutter, only summer months are shown. These are the months when sustained warm nighttime temperatures are most likely to occur. The two vertical lines mark the reported heat wave.}
 \label{fig:TSAM}
\end{figure}

Figure~\ref{fig:TSAM} follows the same display format as Figure~\ref{fig:TS}, but with observed 2~AM temperatures as the response. Because the predictor is the projected 2~PM temperature at the $\tau = 0.90$ quantile, this upper-tail structure is carried forward 12 hours into the nighttime period. The fitted quantile regression smoother preserves the temporal shape induced by persistent warm conditions, while reflecting the cooler overall level of nocturnal temperatures. Despite this diurnal shift, the fitted values track the observed warmer 2~AM temperatures quite closely, including temperatures at or above the 90th percentile.

\section{True Forecasting for the Holdout Data from 2021} 

The conclusions from Figure~\ref{fig:TS} and Figure~\ref{fig:TSAM} depend on fitted values from the training data. Fitted values are not forecasts even though the predictors from the weather station data were lagged by 14 days. There is some evidence that forecasting skill is likely to be high, but credible holdout samples are required for ``honest'' forecasting and proper empirical estimates of forecast uncertainty.

A variant on the common split sample method for adaptive conformal prediction regions was used \citep{Romano2019, Hyndman2021}. Recall that for both the 2~PM and 2~AM response variables, there are data for years 2019, 2020, and 2021. The data from 2020 were used for training the quantile gradient boosting algorithm. The data from 2021 are a pristine holdout sample with which legitimate forecasting  can be undertaken. The data from 2019 are a pristine holdout sample that can be used to calibrate nonconformal scores essential for estimating forecasting uncertainty. All three datasets can be seen as realized from the same joint probability distribution. Still, one should require empirical support for that data generation claim.

\begin{figure}[t]
\centering
  \includegraphics[width=0.9\textwidth]{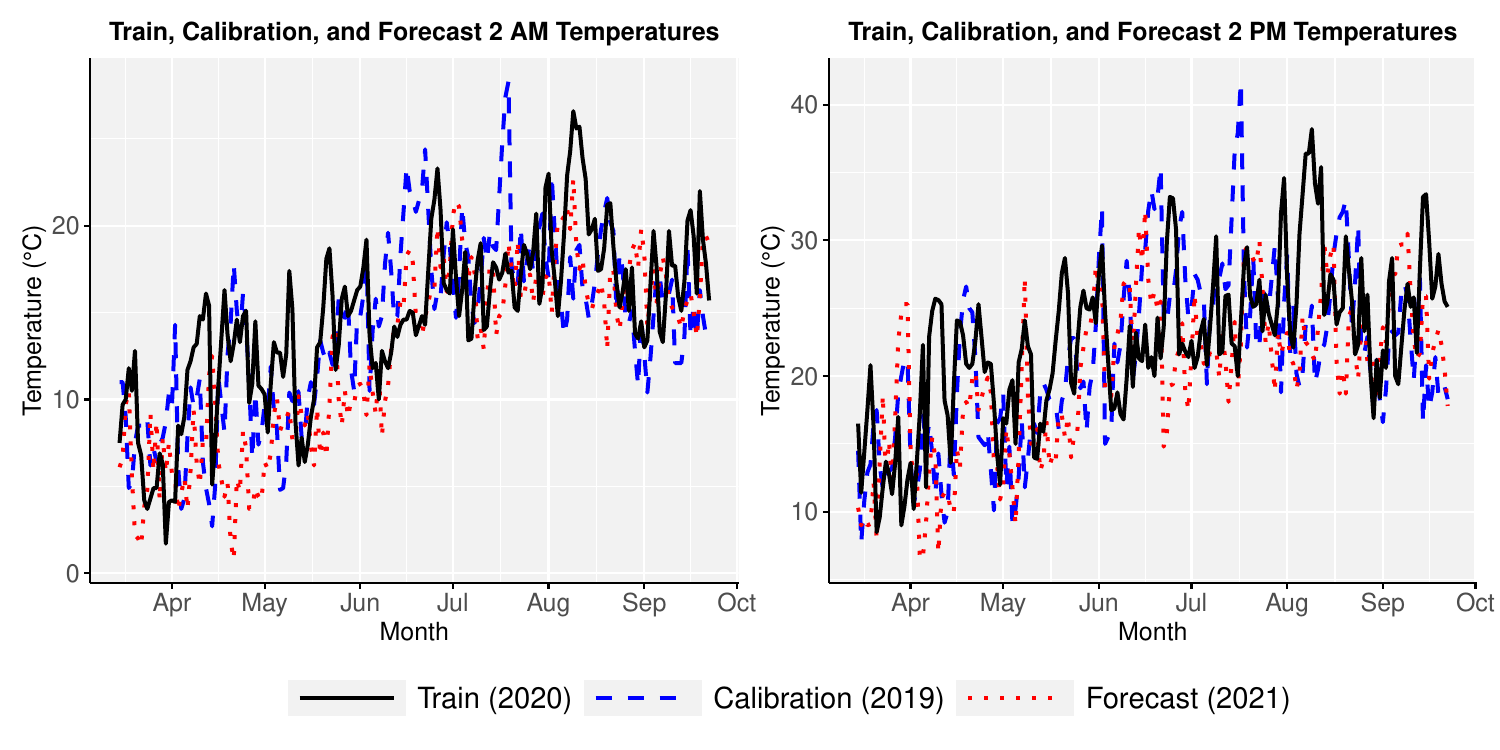}
  \caption{For the 2019, 2020, and 2021 datasets, the left panel displays the 2~AM temperatures. The right panel displays the 2~PM temperatures. For both, the day counter serves as the x-axis, running from early spring to late summer. A legend at the bottom identifies each dataset by line style.}
 \label{fig:samples3}
\end{figure}

Figure~\ref{fig:samples3} provides a visual assessment of comparability across the training, calibration, and forecasting datasets. For both 2~AM and 2~PM temperatures, 
all three series exhibit nearly identical seasonal evolution, suggesting that they are governed by the same large-scale radiative and synoptic forcing. Superimposed on these relatively smooth seasonal trends are intermittent peaks and valleys. Some of these align across datasets, while others do not. There is no physical reason why high temperature peaks and low temperature valleys should occur on the exact same dates in each of the three years. Likewise, the weather station predictors and unmeasured weather features will not have the same values on the same dates across the three years. Boundary layer dynamics effectively preclude temporal matches in peaks and valleys year to year because turbulent convective instabilities are chaotic at short timescales.

In short, large-scale dynamics are shared, but small-scale turbulence is not. At least provisionally, proceeding with a form of statistical comparability leads to some optimism that forecasting uncertainty can be properly assessed. This premise is examined through a comparison between the nominal conformal coverage and the empirical coverage achieved in the 2021 holdout data. 

\subsection{Temporal Dependence and Exchangeability}

Separate calibration residuals were computed for the 2~PM and 2~AM responses by applying the respective fitted algorithms to the 2019 holdout data via \texttt{predict()}. The two sets of residuals are candidate nonconformal scores. However, both exhibited substantial temporal dependence. Such dependence violates the exchangeability assumption required for conformal inference and must be addressed prior to constructing prediction regions \citep{Chernozhukov2018}. The approach employed here is a simple and transparent AR(1) correction with well-established diagnostics.

An AR(1) model was fit separately to each set of candidate nonconformal scores.The resulting residuals from both AR(1) applications, often referred to as innovations, were empirically indistinguishable from white noise. No remaining temporal dependence was evident in autocorrelation functions, and Ljung--Box tests failed to reject the null hypothesis of no serial correlation.

The AR(1) corrections are treated as extensions of their respective training algorithms rather than as post hoc adjustments, which preserves the formal validity of the conformal coverage guarantee. Conformal inference treats the training procedure as given. Some training procedures will perform better than others, and precision can be affected. But the finite sample, coverage claims for conformal inference remain valid as long as exchangeability holds for the nonconformal scores.

White-noise innovations may be regarded as approximately exchangeable and can, therefore, serve in practice as valid nonconformal scores. To construct adaptive conformal prediction regions \citep{Romano2019}, conditional quantile algorithms were fit separately to the two innovation series. Initial attempts using quantile gradient boosting proved unstable for both the 2~PM and 2~AM nonconformal scores, with little improvement beyond the first boosting iteration. A Quantile regression forest was substituted and yielded stable fits in both settings.\footnote
{
Quantile gradient boosting directly minimizes a global quantile loss to
estimate conditional quantile functions \citep{Ridgeway2024}, whereas
quantile regression forests construct trees using variance-based splits and recover quantiles through post hoc evaluation of empirical response distributions within terminal nodes \citep{Meinshausen2006}. When targeting extreme quantiles, loss-based optimization can become unstable due to data sparsity, whereas quantile regression forests are less sensitive to data sparsity because quantile estimation is deferred until after forest construction. Because the 2~PM and 2~AM innovation series exhibit no detectable temporal dependence, the resampling inherent in quantile regression forests does not induce distortions associated with serial correlation.
}

\subsection{Adaptive Conformal Prediction Regions Using the 2021 Forecasting Data}

Before the forecasting procedures are applied, a coverage probability must be specified for the adaptive conformal prediction regions. Values between 0.60 and 0.95 usually are reasonable. The coverage probability chosen is not a statistically driven decision. In practice, it is guided by anticipating the level of certainty that decision makers will prefer. 

The coverage probability is conventionally expressed as $1-\alpha$, where $\alpha$ is the miscoverage rate. For now, the coverage is specified at 0.80 so that $\alpha= 0.20$. But there are tradeoffs between the coverage probability and forecasting precision. These tradeoffs are examined below.

The first priority was to revisit the use of the 2019 and 2021 weather station data as appropriate holdout samples. One instructive performance criterion is whether the theoretical coverage probability of \emph{at least} $0.80$ is consistent with empirically estimated coverage probabilities for the 2021 forecasting data. For the 2~PM observations, the empirical coverage probability is $0.79$. For the 2~AM observations, the empirical coverage probability is $0.87$. For former is almost exactly on target while the latter is a bit conservative. Both seem consistent with calibration and forecasting datasets realized in the same manner from the same joint probability distribution.

Precision was a second consideration. Table~\ref{tab:precision} shows some summary statistics for the lengths of the 2~PM and 2~AM prediction regions. Summaries are needed because the prediction regions are adaptive. Given the coverage probability, smaller prediction region lengths represent greater precision.

Consistent with adaptive conformal prediction regions, precision varies substantially. For the 2~PM forecasts, the range is about $16^{\circ}\mathrm{C}$, and the mean precision is a little over $8^{\circ}\mathrm{C}$ or about $ \pm 4^{\circ}\mathrm{C}$ around the forecast.  For the 2~AM forecasts, the range is about $6^{\circ}\mathrm{C}$, and the mean precision is a little more than $6^{\circ}\mathrm{C}$ or about $ \pm 3^{\circ}\mathrm{C}$ around the forecast. The diurnal precisions are less varied. 

Some may judge these results to be disappointing, because average precision and the range of the precision are relatively large. However, precision is affected by the choice of the fitting quantile in a quantile supervised learning procedure. Here, the fitted values were obtained using $Q(0.90)$, which by design lies somewhat above the bulk of the data. Because for this analysis, construction of the nonconformal scores begins with the residuals, precision would be substantially improved if $Q(0.50)$ were used instead. But then rare and extreme high temperature would not be the estimation target; \emph{typical} temperatures would be the estimation target.

\begin{table}[t]
\caption{Summary statistics for adaptive conformal prediction region lengths for the 2021 weather station 2~PM and 2~AM data. Smaller values indicate greater forecasting precision.}
\label{tab:precision}
\begin{center}
\begin{tabular}{|cccccc|}
\hline 
Time & Minimum & Q1 & Mean & Q3 & Maximum \\
\hline
2~PM & 4.4 & 6.1 & 8.1 & 9.7 & 20.1 \\
2~AM & 3.9 & 5.5 & 6.5 & 7.3 & 10.5 \\
\hline 
\end{tabular}
\end{center}
\end{table}

If stakeholders find the achieved precision unsatisfactory, the tradeoff between coverage and precision might help. Greater precision can be obtained in exchange for lower coverage; $1-\alpha$ can be viewed as a special kind of tuning parameter. There is, however, a statistical complication if a coverage probability is specified after the data analysis has begun. In that case, post--model-selection inference must be implemented \citep{Sarkar2023}, which adds another layer of complexity.\footnote
{
As a rough approximation, if for this analysis coverage were reduced to $0.70$, the mean length of the prediction region would be reduced by about one quarter.
}

Perhaps an alteration in how conformal coverage is defined and reported can better reflect how forecasts might be used in practice by local authorities. In particular, a local decision-maker might face choices about what actions to take if excessive heat is forecasted to arrive in approximately two weeks. Possible actions might include public service announcements regarding impending heat, visits by nurses to the residences of older or medically vulnerable individuals, and arranging for appropriate staffing and medical supplies in hospital emergency rooms in anticipation of increased incidence of hyperthermia. Most interventions, however, involve costs as well as benefits. Some actions, such as subsidizing residential air conditioning, entail substantial monetary costs and are essentially irreversible. Other interventions, such as home visits by nurses, may be perceived by some as invasions of privacy and carry high opportunity costs because nursing resources usefully could be deployed elsewhere.

A full discussion of decision making by local authorities is well beyond the scope of this paper. Nevertheless, it is useful to outline a simple decision framework that relies \emph{only on information available on the day the forecast is issued}. The basic idea is that when higher temperatures are forecast, more consequential measures may be warranted.

Suppose a small set of $J$ \emph{a priori} temperature thresholds can be determined, based on medical evidence, scientific judgment, and cost tradeoffs, such as $\theta_1 < \theta_2 < \theta_3$; higher temperature thresholds correspond to more consequential interventions. An example might be public service announcements $<$ mandatory water breaks for outdoor workers $<$ increases in hospital staffing. For each day $t$, the decision-maker observes the point forecast and an associated \emph{one-sided} adaptive conformal prediction region lower bound $L_t$. The unknown temperature 14 days in the future is denoted by $Y_t$. An operational interpretation follows directly. For any pre-specified temperature threshold $\theta_j$, if the rule ``act when $L_t \ge \theta_j$'' is used, then on that day when action is taken carries the one-sided risk guarantee
\[
\Pr(Y_t \ge \theta_j) \ge 1-\alpha/2.
\]
This guarantee addresses \emph{decision risk} rather than forecast accuracy. The decision-maker can compare different thresholds by their policy tradeoffs under \emph{the same coverage probability}, using no information beyond that which is observable on the day a forecast is made.

It is important to emphasize what this guarantee does and does not imply. A larger lower bound $L_t$ does not correspond to a higher probability of exceeding the associated threshold. For any pre-specified $\theta_j$, once the condition $L_t \ge \theta_j$ holds, the probability that the future temperature exceeds $\theta_j$ is at least $1-\alpha$, regardless of the numerical value of $\theta_j$.

For some readers, the use of three or more thresholds may raise concerns about cherry-picked statistical results. However, for any given day, the thresholds within the outlined decision-making framework are specified before a temperature forecast and an adaptive conformal prediction region are known. Moreover, for each threshold, the same decision rule applies: take the associated actions $j$ on day $t$ if and only if $L_t \ge \theta_j$. Because the conformal coverage level $\alpha$ is fixed, each rule carries the same unalterable probability guarantee. There is, therefore, no opportunity for p-hacking or for exploiting differences in uncertainty estimates across thresholds.

For a one-sided adaptive conformal prediction region, precision can be represented by the difference in degrees between the region's lower bound and the forecasted temperature. But, the formulation also precludes selecting a preferred threshold based on the precision of an associated prediction region. Although such an approach might be reasonable in other contexts, it fails here. For any given day, there is one forecast, one prediction region lower bound, and one precision regardless of the temperature threshold specified. Any $\theta_j$ and $L_t$ are both in the same temperature units that can be compared to determine their order. Precision also is in the same temperature units but represents a prediction region length that properly cannot be compared to any $\theta_j$ temperature. Suppose, for instance, the precision of the prediction region length on a given day is $4^{\circ}\mathrm{C}$. How does one order that length with respect to, say, $\theta_j = 24^{\circ}\mathrm{C}$? They measure very different things.

Suppose $\theta_j$ is specified such that it just happens to correspond to the 90th percentile. Note that $\theta_j$ has no relationship to the coverage probability. Some temperatures below $\theta_j$ can be unpleasant, but are not treated as a substantial threat to public or ecosystem health because they fall below $\theta_j$.

\begin{figure}[h!]
\centering
  \includegraphics[width=0.9\textwidth]{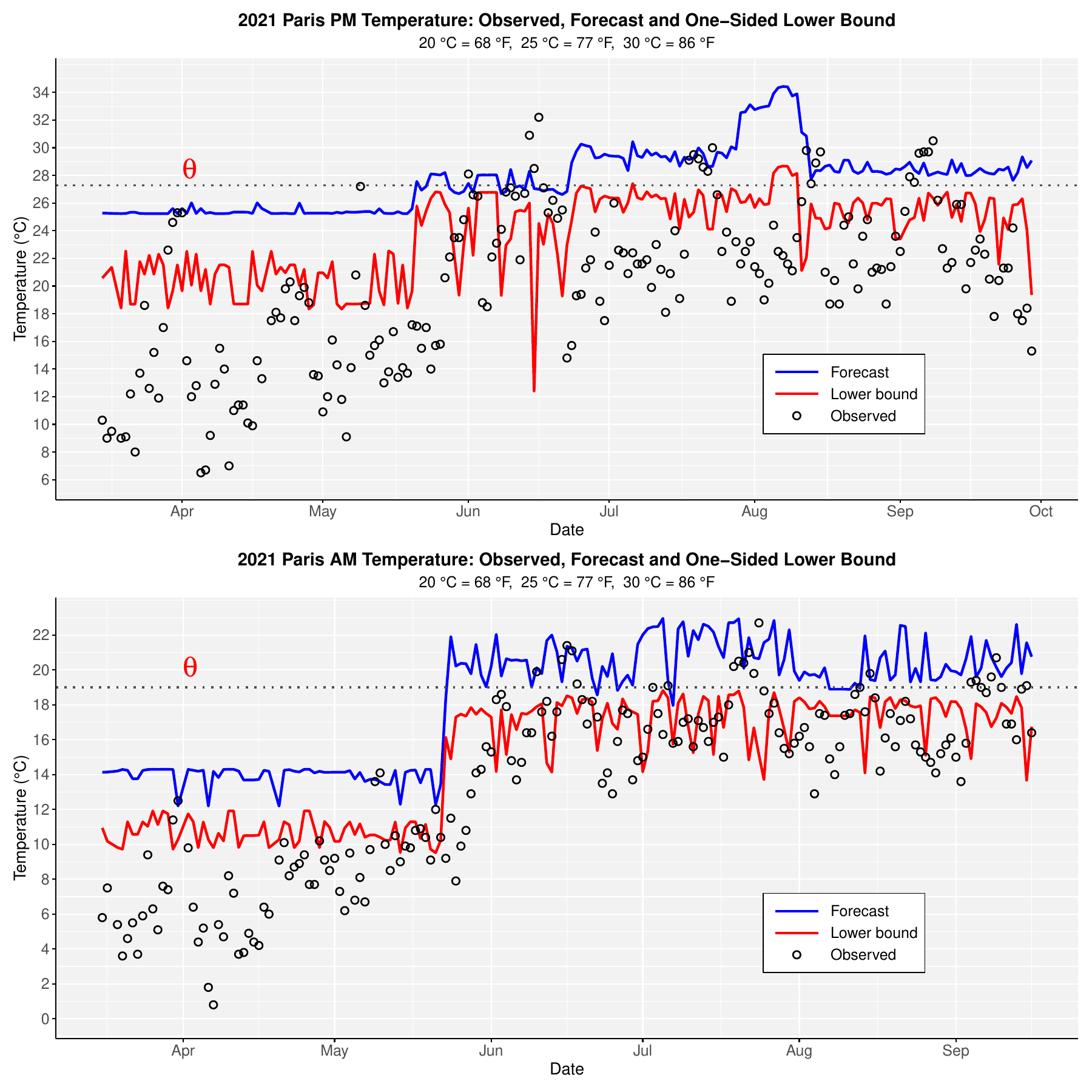}
  \caption{The top panel displays the 2~PM temperatures and the vertical axis, and the bottom panel displays the 2~AM temperatures on the vertical axis. For both, months are on the horizontal axis, the open circles are the observed temperatures, the red line is the adaptive conformal prediction region lower bounds, the blue line is the forecasts and the horizontal dotted line specifies the value of $\theta_j$. A legend at the bottom identifies each plotting symbol."}
 \label{fig:conformal}
\end{figure}

Figure~\ref{fig:conformal} illustrates a simple decision-making example. In practice decisions would be made day by day. The figure is meant to represent how well the day by day decisions would perform over the warm weather months.
Consider first the top panel for the 2~PM data. The open circles represent actual future temperatures. These temperatures would not be known when real decisions would need to be made, but are available in the 2021 forecasting data and help demonstrate how the decision framework performs if implemented as designed.

 Suppose $\theta_j$ is specified such that it just happens to correspond to the 90th percentile. Note that $\theta_j$ has no relationship to the coverage probability. Some temperatures below $\theta_j$ can be unpleasant, but are not treated as a substantial threat to public or ecosystem health because they fall below $\theta_j$.
 
The blue line shows the forecasts. Much as with the fit of the training data, the forecasts fall above most of the observed future temperatures and sometimes above the temperature threshold $\theta_j$. If a forecast is at or above $\theta_j$, some policy response could be authorized. 

But are such forecasts to be believed? The red line shows for each day the $L_t$ values, which are the lower bounds from the adaptive conformal prediction regions. Their one-sided coverage probability is $0.90$. The forecasts are all above $L_t$ by construction.

In Figure~\ref{fig:conformal}, all of the (unknown) future temperatures at or above the threshold $\theta_j$ also fall above their respective lower bound $L_t$. They are all, therefore, true positives. For this example, the forecasts provide good guidance because $\theta_j$ happens to fall at the quantile that served as the estimation target for the quantile gradient boosting. If $\theta_j$ corresponded to lower quantile, performance would decline, perhaps even to a disastrous level. For example, if $\theta_j$ were near the median, most of the forecasts would be false positives. But if future temperatures at or above the median were a major concern for public or ecosystem health, the value of $\tau$ for the quantile gradient boosting should have been set at $0.50$. Then forecasting performance probably would be excellent. In other words, $\tau$ for 2~PM temperatures should be set with the quantile of $\theta_j$ in mind when the algorithm is trained. Possible programmatic responses to excessive heat should be considered in some detail before training the quantile gradient boosting algorithm.

The same reasoning applies to the bottom panel in Figure~\ref{fig:conformal} in which 2~AM panel is a direct extension of the same framework, with $\tau$ chosen to match the nocturnal policy threshold, The quantile gradient boosting would be trained as before with its appropriate value of $\tau$ for the 2~PM data. But when the quantile regression smoother was applied to the 2~AM data, the value of $\tau$ was chosen with respect to the nocturnal value of $\theta_j$ and its quantile. For expositional ease, $\theta_j$ for the 2~AM data in the lower panel is assumed to fall at Q(0.90) for the 2~AM temperatures.

\FloatBarrier

\section{Discussion}

The years chosen for the training, calibration, and forecasting data were determined before the data were curated and analyzed. No effort was made to anticipate how the analysis would unfold. The results are not compromised by cherry-picked data.

Nevertheless, how well the analysis and forecasting procedures work must depend on the local climate. Paris lies near to the Loire Valley. It has a temperate oceanic climate coupled with urban heat island effects. The winters are mild and the summers are warm. Cloud cover is common, and humidity is moderate. Rain falls evenly throughout the year. There are many areas around the globe that properly could be described in a similar manner. 

However, there are locales where the climate is very different such as the American Southwest, sites near or above the Arctic Circle, and locations subject to monsoons. 
Even if the methods applied here perform correctly, the empirical results could differ substantially in other climates. For example,  Cairo and Los Angeles can experience very rapid temperature changes when onshore winds bringing moist, cool air are reversed within an hour or two by dry desert winds \citep{Warner2004}. In another extreme case, Svalbard, Norway has no nocturnal conditions at all from approximately late April to late August; there is in principle solar radiation 24 hours a day for nearly 100 days. Consequently, the usual boundary layer decoupling can effectively disappear with consequence for temperature inversions \citep{Oke1987, Peng2023,Jozef2024}.

Even in a single location, there can be significant variation from year to year because of El Ni\~{n}o events, displacements of the Jet Stream, and other large scale processes affecting slow moving seasonal cycles. The local physics is unchanged, but the states and values of weather features can vary in important ways. More subtle are changes in rapid weather dynamics when nonlinear, small-scale processes accumulate as stochastic noise. 

In short, training data, calibration data, and forecasting data differ significantly across climates and can also vary substantially from year to year in the same locale. If the lagged predictors do not change in ways that are associated with such variability, forecasting skill and uncertainty quantification will suffer. Whether these problems are common is an empirical question. Weather station data are available at thousands of sites around the globe. Replications are straightforward to mount. 

The choice of $Q(0.90)$ as the operational definition of rare and excessive heat should be informed by local factors as well as statistical properties.  Architecture, technology, and culture affect the science and the policy. Urban heat islands are one illustration. Afternoon siestas are another. Fragile public or ecosystem health might argue for a lower quantile. If effective adaptive practices already exist, a higher quantile might be appropriate. 

With a 14~day lead time, a range of proactive measures could be implemented or at least better planned \citep{David2015}. Examples include radio and TV announcements providing information on symptoms of heat-related illnesses; preparing residences for excessive heat; outreach to vulnerable groups such as elderly individuals living alone; preparing public cooling buildings that can be used as refuges; providing proper staffing and provisioning of hospital emergency rooms and paramedic vehicles; adjusting work schedules and mandating water breaks during excessive heat; eliminating or minimizing outdoor activities for schoolchildren; utility coordination anticipating higher electricity use; watering vulnerable plants, shrubs, and trees; making cool water available to pets and zoo animals; and having firefighters and their supporting equipment moved near undeveloped land at risk from wildfires.

\section{Conclusions}

The results suggest that rare and excessive temperatures can be forecast two weeks in advance, at least in temperate oceanic climates similar to Paris. Nocturnal as well as diurnal temperatures are forecast with promising accuracy and valid estimates of uncertainty. A single weather station can provide appropriate spatial resolution for an urban warning system and subsequent planning. Analyses can be undertaken on a laptop or desktop computer equipped with Python or R with modest computational requirements. Pseudocode is provided in the appendix.

\clearpage
%%%%%%%%%%%%%%%%%%%%%%%%%%%%%%%%%%%%%%%%%%%%%%%%%%%%%%%%%%%%%%%%%%%%%
\section*{Data Availability Statement}
The meteorological data used in this study are publicly available from the sources cited in the manuscript. No proprietary or confidential data were used. The processed datasets and analysis code will be made available upon reasonable request and/or in a public repository upon acceptance.

%%%%%%%%%%%%%%%%%%%%%%%%%%%%%%%%%%%%%%%%%%%%%%%%%%%%%%%%%%%%%%%%%%%%%
% APPENDIX
% AMS uses \appendix (or \appendix[A], \appendix[B], etc. for multiple)
% Note: appendix sections start with \subsection, NOT \section
%%%%%%%%%%%%%%%%%%%%%%%%%%%%%%%%%%%%%%%%%%%%%%%%%%%%%%%%%%%%%%%%%%%%%

\section*{Appendix: Pseudocode}
\subsection*{Pseudocode 1: 2~PM Temperature Forecasts}

\begin{algorithmic}[1]

\State \textbf{Step 1 (Input and data split).}
This procedure applies to 2~PM temperatures; the analogous 2~AM procedure appears in Pseudocode~2. Let \(D_1, D_2, D_3\) denote the datasets for training (March--September 2020), calibration (2019), and forecasting (2021), respectively.
For each dataset \(D_k\), let \(X^{(k)}_{t-14}\) denote the vector of 14-day lagged predictor values, and let \(y^{pm,(k)}_t\) denote the observed 2~PM temperature.
Predictors have identical definitions across datasets, but their realizations differ year to year.

Let \(\tau_0\) denote the quantile level used for point prediction (e.g., \(\tau_0 = 0.80\)), and let \(\alpha\) determine the desired coverage probability \(1-\alpha\).

\State \textbf{Step 2 (Train base algorithm on \(D_1\)).}
Using quantile level \(\tau_0\), train a boosting algorithm \(B\) on \(D_1\) to estimate the conditional \(\tau_0\)-quantile of \(y^{pm,(1)}_t\) given \(X^{(1)}_{t-14}\)
(e.g., quantile gradient boosting). Denote the trained base algorithm by \(\hat{B}\).

\State \textbf{Step 3 (Apply \(\hat{B}\) to calibration data \(D_2\)).} For each calibration time \(t = 1,\ldots,T_2\), compute the fitted value
\[
w^{(2)}_t = \hat{B}\!\left(X^{(2)}_{t-14}\right),
\]
with corresponding observed 2~PM temperature \(y^{pm,(2)}_t\).

\State \textbf{Step 4 (Compute calibration residuals).} For each \(t = 1,\ldots,T_2\), compute the residual
\[
r_t = y^{pm,(2)}_t - w^{(2)}_t .
\]

\State \textbf{Step 5 (Whiten calibration residuals).} Fit an AR(1) time-series model to the residual sequence \(\{r_t\}_{t=1}^{T_2}\) and extract the innovations
\[
z_t, \qquad t = 1,\ldots,T_2,
\]
which are treated as the nonconformal scores.

\State \textbf{Step 6 (Train score algorithm on calibration data).} Using the calibration pairs \(\{(w^{(2)}_t, z_t) : t = 1,\ldots,T_2\}\), train a score algorithm \(Q\) (e.g., a quantile random forest) to estimate conditional quantiles of \(z_t\) given the fitted value \(w^{(2)}_t\). Denote the trained score algorithm by \(\hat{Q}\), and let \(\hat q_{\gamma}(w)\) denote the fitted conditional \(\gamma\)-quantile of \(z \mid w\).

\State \textbf{Step 7 (Point forecasts for \(D_3\)).} For each forecasting time \(t\) in \(D_3\), compute the 2~PM point forecast
\[
w^{for}_t = \hat{B}\!\left(X^{(3)}_{t-14}\right).
\]

\State \textbf{Step 8 (Adaptive score quantiles).} For each forecasting time \(t\), evaluate the fitted conditional score quantiles at \(w^{for}_t\):
\[
q_{t,\alpha/2} = \hat q_{\alpha/2}\!\left(w^{for}_t\right), \qquad q_{t,1-\alpha/2} = \hat q_{1-\alpha/2}\!\left(w^{for}_t\right).
\]

\State \textbf{Step 9 (Construct adaptive conformal prediction interval).} Form the adaptive conformal prediction interval for the 2~PM temperature at time \(t\):
\[
\bigl[w^{for}_t + q_{t,\alpha/2},\; w^{for}_t + q_{t,1-\alpha/2}\bigr].
\]

\State \textbf{Step 10 (Output).}
For each forecasting case in \(D_3\), report the 2~PM point forecast \(w^{for}_t\)
and the corresponding adaptive conformal prediction interval from Step~9.

\end{algorithmic}

\subsection*{Pseudocode 2: 2~AM Temperature Forecasts}

\begin{algorithmic}[1]

\State \textbf{Step 1 (Input and data split).}
This procedure constructs adaptive prediction intervals for 2~AM temperatures using the 2~PM forecasts from Pseudocode~1. Let \(D_1, D_2, D_3\) denote the datasets for training (March--September 2020), calibration (2019), and forecasting (2021), respectively.

For each dataset \(D_k\), let \(X^{(k)}_{t-14}\) denote the vector of 14-day lagged predictor values, let \(y^{pm,(k)}_t\) denote the observed 2~PM temperature, and let \(y^{am,(k)}_t\) denote the observed 2~AM temperature. From Pseudocode~1, take the trained 2~PM prediction algorithm \(\hat{B}\). Let \(\alpha\) determine the desired coverage probability \(1-\alpha\).

\State \textbf{Step 2 (2~PM fitted values on \(D_1\) and \(D_2\)).} For each year \(k \in \{1,2\}\) and each time \(t\) in \(D_k\), compute the 2~PM fitted values
\[
w^{(k)}_t = \hat{B}\!\left(X^{(k)}_{t-14}\right).
\]

\State \textbf{Step 3 (Thermal--inertia predictor for 2~AM).} For each \(k \in \{1,2\}\) and for indices \(t\) where a previous-day 2~PM fit is available, define the one-day--lagged 2~PM predictor
\[
u^{(k)}_t = w^{(k)}_{t-1},
\]
which represents the previous-day 2~PM fitted temperature driving the next 2~AM temperature via thermal inertia.

\State \textbf{Step 4 (Fit 2~AM quantile smoother on \(D_1\)).} Using the training-year pairs \(\{(u^{(1)}_t, y^{am,(1)}_t)\}\) for all admissible \(t\), fit a quantile smoothing spline targeting the \(\tau = 0.90\) conditional quantile of the 2~AM temperature given the lagged 2~PM fitted value. Denote the trained 2~AM quantile smoother by \(\hat{L}\).

\State \textbf{Step 5 (2~AM fitted values on calibration data \(D_2\)).} For each admissible time \(t\) in \(D_2\), compute the fitted 2~AM baseline quantile
\[
m^{(2)}_t = \hat{L}\!\left(u^{(2)}_t\right),
\]
with corresponding observed 2~AM temperature \(y^{am,(2)}_t\).

\State \textbf{Step 6 (Calibration residuals).} For each admissible calibration time \(t\), compute the 2~AM residual
\[
r^{am}_t = y^{am,(2)}_t - m^{(2)}_t .
\]

\State \textbf{Step 7 (Whiten calibration residuals).} Fit an AR(1) time-series model to the residual sequence \(\{r^{am}_t\}\) and extract the resulting innovations
\[
z^{am}_t,
\]
which are treated as the nonconformal scores for 2~AM.

\State \textbf{Step 8 (Train 2~AM score algorithm on calibration data).} Using the calibration pairs \(\{(m^{(2)}_t, z^{am}_t)\}\), train a 2~AM score algorithm \(Q^{am}\) (e.g., a quantile regression forest) to estimate conditional quantiles of \(z^{am}_t\) given \(m^{(2)}_t\). Denote the trained score algorithm by \(\hat{Q}^{am}\),
and let \(\hat q^{am}_{\gamma}(m)\) denote the fitted conditional \(\gamma\)-quantile of \(z^{am} \mid m\).

\State \textbf{Step 9 (Point forecasts for 2~AM on \(D_3\)).} For each forecasting time \(t\) in \(D_3\), first compute the 2~PM point forecast
\[
w^{for}_t = \hat{B}\!\left(X^{(3)}_{t-14}\right),
\]
then define the corresponding thermal--inertia predictor
\[
u^{for}_t = w^{for}_{t-1},
\]
whenever the previous-day 2~PM forecast is available, and obtain the 2~AM baseline forecast
\[
m^{for}_t = \hat{L}\!\left(u^{for}_t\right).
\]

\State \textbf{Step 10 (Adaptive score quantiles for 2~AM).} For each forecasting time \(t\) with baseline forecast \(m^{for}_t\), evaluate the fitted conditional score quantiles:
\[
q^{am}_{t,\alpha/2} = \hat q^{am}_{\alpha/2}\!\left(m^{for}_t\right), \qquad
q^{am}_{t,1-\alpha/2} = \hat q^{am}_{1-\alpha/2}\!\left(m^{for}_t\right).
\]

\State \textbf{Step 11 (Construct prediction intervals and output).}
For each forecasting case in \(D_3\), form the adaptive conformal prediction interval for the 2~AM temperature at time \(t\) as
\[
\bigl[m^{for}_t + q^{am}_{t,\alpha/2},\;
      m^{for}_t + q^{am}_{t,1-\alpha/2}\bigr],
\]
and report the 2~AM baseline forecast \(m^{for}_t\) together with this interval.

\end{algorithmic}

%%%%%%%%%%%%%%%%%%%%%%%%%%%%%%%%%%%%%%%%%%%%%%%%%%%%%%%%%%%%%%%%%%%%%
% REFERENCES
%%%%%%%%%%%%%%%%%%%%%%%%%%%%%%%%%%%%%%%%%%%%%%%%%%%%%%%%%%%%%%%%%%%%%
\bibliographystyle{plainnat}
\bibliography{DayNight}

\begin{thebibliography}{55}
\providecommand{\natexlab}[1]{#1}
\providecommand{\url}[1]{\texttt{#1}}
\expandafter\ifx\csname urlstyle\endcsname\relax
  \providecommand{\doi}[1]{doi: #1}\else
  \providecommand{\doi}{doi: \begingroup \urlstyle{rm}\Url}\fi

\bibitem[Anderson and Bell(2009)]{Anderson2009}
G.~B. Anderson and M.~L. Bell.
\newblock Weather-related mortality: How heat, cold, and heat waves affect
  mortality in the united states.
\newblock \emph{Epidemiology}, 20\penalty0 (2):\penalty0 205--213, 2009.
\newblock \doi{10.1097/EDE.0b013e318190ee08}.

\bibitem[Angelopoulos et~al.(2024)Angelopoulos, Barber, and
  Bates]{Angelopoulos2025}
A.~N. Angelopoulos, R.~F. Barber, and S.~Bates.
\newblock Theoretical foundations of conformal prediction, 2024.
\newblock URL \url{https://arxiv.org/abs/2411.11824}.
\newblock Preprint; forthcoming with Cambridge University Press.

\bibitem[Ballester et~al.(2023)Ballester, Quijal-Zamorano, and
  M{\'e}ndez-Turrubiates]{Ballester2023}
J.~Ballester, M.~Quijal-Zamorano, and R.F. et~al. M{\'e}ndez-Turrubiates.
\newblock Heat-related mortality in europe during the summer of 2022.
\newblock \emph{Nature Medicine}, 29:\penalty0 1857--1866, 2023.
\newblock \doi{10.1038/s41591-023-02419-z}.
\newblock URL \url{https://doi.org/10.1038/s41591-023-02419-z}.

\bibitem[Bodnar et~al.(2025)Bodnar, Bruinsma, Lucic, et~al.]{Bodnar2025}
C.~Bodnar, W.~P. Bruinsma, A.~Lucic, et~al.
\newblock A foundation model for the earth system.
\newblock \emph{Nature}, 641:\penalty0 1180--1187, 2025.
\newblock \doi{10.1038/s41586-025-09005-y}.

\bibitem[Breiman(2001)]{Breiman2001}
L.~Breiman.
\newblock Statistical modeling: The two cultures.
\newblock \emph{Statistical Science}, 16\penalty0 (3):\penalty0 199--231, 2001.
\newblock \doi{10.1214/ss/1009213726}.
\newblock URL \url{https://doi.org/10.1214/ss/1009213726}.

\bibitem[Breshears et~al.(2021)Breshears, Fontaine, Ruthrof, Field, Feng,
  Burger, Law, Kala, and Hardy]{Breshears2021}
D.~Breshears, J.~Fontaine, K.~Ruthrof, J.~Field, X.~Feng, J.~Burger, D.~Law,
  J.~Kala, and G.~Hardy.
\newblock Underappreciated plant vulnerabilities to heat waves.
\newblock \emph{New Phytologist}, 231:\penalty0 32--39, 2021.
\newblock \doi{10.1111/nph.17348}.
\newblock URL \url{https://doi.org/10.1111/nph.17348}.

\bibitem[Chernozhukov et~al.(2018)Chernozhukov, W\"{u}thrich, and
  Zhu]{Chernozhukov2018}
V.~Chernozhukov, K.~W\"{u}thrich, and Y.~Zhu.
\newblock Exact and robust conformal inference methods for predictive machine
  learning with dependent data.
\newblock In S.~Bubeck, V.~Perchet, and P.~Rigollet, editors, \emph{Proceedings
  of the 31st Conference On Learning Theory}, volume~75 of \emph{Proceedings of
  Machine Learning Research}, pages 732--749. PMLR, 06--09 Jul 2018.
\newblock URL \url{https://proceedings.mlr.press/v75/chernozhukov18a.html}.

\bibitem[Cvijanovic et~al.(2023)Cvijanovic, Mistry, Begg, Gasparrini, and
  Rod\'{o}]{Cvijanovic2023}
I.~Cvijanovic, M.N. Mistry, J.D. Begg, A.~Gasparrini, and X.~Rod\'{o}.
\newblock Importance of humidity for characterization and communication of
  dangerous heatwave conditions.
\newblock \emph{npj Climate and Atmospheric Science}, 6:\penalty0 33, 2023.
\newblock \doi{10.1038/s41612-023-00346-x}.
\newblock URL \url{https://doi.org/10.1038/s41612-023-00346-x}.

\bibitem[David(2015)]{David2015}
F.~David.
\newblock Pr{\'e}vention des risques li{\'e}s {\`a} la canicule et aux fortes
  chaleurs.
\newblock \emph{La Sant{\'e} en Actions}, 432:\penalty0 33--34, 2015.
\newblock \doi{https://doi.org/10.1038/s43017-022-00371-z}.

\bibitem[Domeisen et~al.(2023)Domeisen, Eltahir, Fischer, Knutti,
  Perkins-Kirkpatrick, Sch{\"a}r, Seneviratne, Weisheimer, and
  Wernli]{Domeisen2023}
D.I.V. Domeisen, E.A.B. Eltahir, E.M. Fischer, R.~Knutti, S.E.
  Perkins-Kirkpatrick, C.~Sch{\"a}r, S.I. Seneviratne, A.~Weisheimer, and
  H.~Wernli.
\newblock Prediction and projection of heatwaves.
\newblock \emph{Nature Reviews Earth \& Environment}, 4:\penalty0 36--50, 2023.
\newblock \doi{https://doi.org/10.1038/s43017-022-00371-z}.

\bibitem[Fischer et~al.(2023)Fischer, Beyerle, Schleussner,
  et~al.]{Fischer2023}
E.~M. Fischer, U.~Beyerle, C.-F. Schleussner, et~al.
\newblock Storylines for unprecedented heatwaves based on ensemble boosting.
\newblock \emph{Nature Communications}, 2023.
\newblock \doi{10.1038/s41467-023-40112-4}.

\bibitem[Friedman(2001)]{Friedman2001}
J.~H. Friedman.
\newblock Greedy function approximation: A gradient boosting machine.
\newblock \emph{The Annals of Statistics}, 29\penalty0 (5):\penalty0
  1189--1232, 2001.
\newblock \doi{10.1214/aos/1013203451}.
\newblock URL \url{https://doi.org/10.1214/aos/1013203451}.

\bibitem[Friedman(2002)]{Friedman2002}
J.~H. Friedman.
\newblock Stochastic gradient boosting.
\newblock \emph{Computational Statistics \& Data Analysis}, 38\penalty0
  (4):\penalty0 367--378, 2002.
\newblock \doi{10.1016/S0167-9473(01)00065-2}.
\newblock URL \url{https://doi.org/10.1016/S0167-9473(01)00065-2}.

\bibitem[Fu(2025)]{Fu2025}
B.~Fu.
\newblock State of the science fact sheet: Uncertainty in forecasting weather
  and water.
\newblock Technical Report 69977, National Oceanic and Atmospheric
  Administration, 2025.
\newblock URL \url{https://doi.org/10.25923/3r83-dt50}.

\bibitem[Gettleman and Rood(2016)]{Gettleman2016}
A.~Gettleman and R.~B. Rood.
\newblock \emph{Demystifying Climate Models: A User's Guide to Earth System
  Models}.
\newblock Springer Praxis Books. Springer, Cham, 2016.
\newblock \doi{10.1007/978-3-662-52838-0}.
\newblock URL \url{https://doi.org/10.1007/978-3-662-52838-0}.

\bibitem[Goodfellow et~al.(2016)Goodfellow, Bengio, and
  Courville]{Goodfellow2017}
I.~Goodfellow, Y.~Bengio, and A.~Courville.
\newblock \emph{Deep Learning}.
\newblock MIT Press, Cambridge, MA, 2016.
\newblock ISBN 0262035618.

\bibitem[Hao et~al.(2022)Hao, Liu, Zhang, Ying, Feng, Su, and Zhu]{Hao2022}
Z.~Hao, S.~Liu, Y.~Zhang, C.~Ying, Y.~Feng, H.~Su, and J.~Zhu.
\newblock Physics-informed machine learning: A survey on problems, methods and
  applications.
\newblock arXiv preprint arXiv:2211.08064, 2022.
\newblock URL \url{https://arxiv.org/abs/2211.08064}.
\newblock Accessed: 28 December 2025.

\bibitem[Hastie et~al.(2009)Hastie, Tibshirani, and Friedman]{Hastie2009}
T.~Hastie, R.~Tibshirani, and J.~Friedman.
\newblock \emph{The Elements of Statistical Learning: Data Mining, Inference,
  and Prediction}.
\newblock Springer, New York, 2 edition, 2009.

\bibitem[He et~al.(2022)He, Kim, Hashizume, et~al.]{He2022}
C.~He, H.~Kim, M.~Hashizume, et~al.
\newblock The effects of night-time warming on mortality burden under future
  climate change scenarios: A modeling study.
\newblock \emph{The Lancet Planetary Health}, 6\penalty0 (8):\penalty0
  e648--e657, 2022.
\newblock \doi{10.1016/S2542-5196(22)00139-5}.
\newblock URL \url{https://doi.org/10.1016/S2542-5196(22)00139-5}.

\bibitem[Hopke(2020)]{Hopke2020}
J.E. Hopke.
\newblock Connecting extreme heat events to climate change: Media coverage of
  heat waves and wildfires.
\newblock \emph{Environmental Communication}, 14\penalty0 (4):\penalty0
  492--508, 2020.
\newblock \doi{10.1080/17524032.2019.1687537}.
\newblock URL \url{https://doi.org/10.1080/17524032.2019.1687537}.

\bibitem[Hulme et~al.(2008)Hulme, Dassai, Lorenzoni, and Nelson]{Hulme2008}
M.~Hulme, S.~Dassai, I.~Lorenzoni, and D.R. Nelson.
\newblock Unstable climates: Exploring the statistical and social constructions
  of 'normal' climate.
\newblock \emph{Geoforum}, 40:\penalty0 197--205, 2008.
\newblock \doi{10.1016/j.geoforum.2008.09.010}.
\newblock URL \url{https://doi.org/10.1016/j.geoforum.2008.09.010}.

\bibitem[Hyndman and Athanasopoulos(2021)]{Hyndman2021}
R.J. Hyndman and G.~Athanasopoulos.
\newblock \emph{Forecasting: Principles and Practice}.
\newblock OTexts, Melbourne, 3 edition, 2021.
\newblock URL \url{https://otexts.com/fpp3/}.

\bibitem[Jozef et~al.(2024)Jozef, Cassano, Dahlke, Dice, Cox, and
  de~Boer]{Jozef2024}
G.C. Jozef, J.J. Cassano, S.~Dahlke, M.~Dice, C.~J. Cox, and G.~de~Boer.
\newblock An overview of the vertical structure of the atmospheric boundary
  layer in the central {Arctic} during {MOSAiC}.
\newblock \emph{Atmospheric Chemistry and Physics}, 24:\penalty0 1429--1450,
  2024.
\newblock \doi{10.5194/acp-24-1429-2024}.

\bibitem[Kearns and Roth(2019)]{Kearns2019}
M.~Kearns and A~Roth.
\newblock \emph{The Ethical Algorithm: The Science of Socially Aware Algorithm
  Design}.
\newblock Oxford University Press, New York, 2019.

\bibitem[Koenker and Machado(1999)]{Koenker1999}
R.~Koenker and J.A.F. Machado.
\newblock Goodness of fit and related inference processes for quantile
  regression.
\newblock \emph{Journal of the American Statistical Association}, 94\penalty0
  (448):\penalty0 1296--1310, 1999.
\newblock \doi{10.1080/01621459.1999.10473882}.
\newblock URL \url{https://doi.org/10.1080/01621459.1999.10473882}.

\bibitem[Koenker et~al.(1994)Koenker, Ng, and Portnoy]{Koenker1994}
R.~Koenker, P.~Ng, and S.~Portnoy.
\newblock Quantile smoothing splines.
\newblock \emph{Biometrika}, 81\penalty0 (4):\penalty0 673--680, 1994.
\newblock \doi{10.1093/biomet/81.4.673}.

\bibitem[Koenker et~al.(2017)Koenker, Chernozhukov, He, and Peng]{Koenker2017}
R.~Koenker, V.~Chernozhukov, X.~He, and L.~Peng, editors.
\newblock \emph{Handbook of Quantile Regression}.
\newblock Chapman and Hall/CRC, Boca Raton, FL, 1 edition, 2017.
\newblock \doi{10.1201/9781315120256}.

\bibitem[Li et~al.(2024)Li, Mann, Wehner, et~al.]{Li2024}
X.~Li, M.E. Mann, M.F. Wehner, et~al.
\newblock Role of atmospheric resonance and land--atmosphere feedbacks as a
  precursor to the june 2021 pacific northwest heat dome event.
\newblock \emph{Proceedings of the National Academy of Sciences}, 121\penalty0
  (4):\penalty0 e2315330121, 2024.
\newblock \doi{10.1073/pnas.2315330121}.
\newblock URL \url{https://doi.org/10.1073/pnas.2315330121}.

\bibitem[Mann et~al.(2018)Mann, Rahmstorf, Kornhuber, and Steinman]{Mann2018}
M.E. Mann, S.~Rahmstorf, K.~Kornhuber, and B.A. Steinman.
\newblock Projected changes in persistent extreme summer weather events: The
  role of quasi-resonant amplification.
\newblock \emph{Science Advances}, 4\penalty0 (10):\penalty0 eaat3272, 2018.
\newblock \doi{10.1126/sciadv.aat3272}.
\newblock URL \url{https://doi.org/10.1126/sciadv.aat3272}.

\bibitem[Masselot et~al.(2023)Masselot, Mistry, Vanoli, Schneider, Lungman,
  Garcia-Leon, and et~al.]{Masselot2023}
P.~Masselot, M.~Mistry, J.~Vanoli, R.~Schneider, T.~Lungman, D.~Garcia-Leon,
  and et~al.
\newblock Excess mortality attributed to heat and cold: A health impact
  assessment study in 854 cities in europe.
\newblock \emph{Lancet: Planetary Health}, 7\penalty0 (4):\penalty0 E271--E281,
  2023.
\newblock \doi{10.1016/S2542-5196(23)00023-2}.
\newblock URL
  \url{https://www.thelancet.com/journals/lanplh/article/PIIS2542-5196(23)00023-2/fulltext3}.

\bibitem[McKinnon and Simpson(2022)]{McKinnon2022}
K.A. McKinnon and I.R. Simpson.
\newblock How unexpected was the 2021 pacific northwest heatwave?
\newblock \emph{Geophysical Research Letters}, 49\penalty0 (18), 2022.
\newblock \doi{10.1029/2022GL100380}.
\newblock URL \url{https://doi.org/10.1029/2022GL100380}.

\bibitem[Meinshausen(2006)]{Meinshausen2006}
N.~Meinshausen.
\newblock Quantile regression forests.
\newblock \emph{Journal of Machine Learning Research}, 7:\penalty0 983--999,
  2006.
\newblock URL \url{http://www.jmlr.org/papers/v7/meinshausen06a.html}.

\bibitem[{National Center for Atmospheric Research}(2025)]{NCAR2025}
{National Center for Atmospheric Research}.
\newblock Determining computational resource needs.
\newblock
  \url{https://ncar-hpc-docs.readthedocs.io/en/latest/allocations/determining-computational-resource-needs/},
  2025.
\newblock Accessed: 31 August 2025.

\bibitem[Oke(1987)]{Oke1987}
T.~R. Oke.
\newblock \emph{Boundary Layer Climates}.
\newblock Routledge, London, 2 edition, 1987.

\bibitem[Pascal et~al.(2021)Pascal, Lagarrigue, Tabai, and et~al.]{Pascal2021}
M.~Pascal, R.~Lagarrigue, A.~Tabai, and et~al.
\newblock Evolving heat waves characteristics challenge heat warning systems
  and prevention plans.
\newblock \emph{International Journal of Biometeorology}, 65:\penalty0
  1683--1694, 2021.
\newblock \doi{10.1007/s00484-021-02123-y}.
\newblock URL \url{https://doi.org/10.1007/s00484-021-02123-y}.

\bibitem[Peng et~al.(2023)Peng, Yang, Shupe, Xi, Han, Chen, Dahlke, and
  Liu]{Peng2023}
S.~Peng, Q.~Yang, M.D. Shupe, X.~Xi, B.~Han, D.~Chen, S.~Dahlke, and C.~Liu.
\newblock The characteristics of atmospheric boundary layer height over the
  {Arctic Ocean} during {MOSAiC}.
\newblock \emph{Atmospheric Chemistry and Physics}, 23:\penalty0 8683--8703,
  2023.
\newblock \doi{10.5194/acp-23-8683-2023}.

\bibitem[Perkins and Alexander(2013)]{Perkins2013}
S.E. Perkins and L.V. Alexander.
\newblock On the measurement of heat waves.
\newblock \emph{Journal of Climate}, 26\penalty0 (13):\penalty0 4500--4517,
  2013.
\newblock \doi{10.1175/JCLI-D-12-00383.1}.
\newblock URL \url{https://doi.org/10.1175/JCLI-D-12-00383.1}.

\bibitem[Petoukhov et~al.(2013)Petoukhov, Rahmstorf, Petri, and
  Schellnhuber]{Petoukhov2013}
V.~Petoukhov, S.~Rahmstorf, S.~Petri, and H.J. Schellnhuber.
\newblock Quasiresonant amplification of planetary waves and recent northern
  hemisphere weather extremes.
\newblock \emph{Proceedings of the National Academy of Sciences}, 110\penalty0
  (14):\penalty0 5336--5341, 2013.
\newblock \doi{10.1073/pnas.1222000110}.
\newblock URL \url{https://doi.org/10.1073/pnas.1222000110}.

\bibitem[Porter(2025)]{Porter2025}
C.~Porter.
\newblock Paris braces for a future of possibly paralyzing heat.
\newblock \emph{The New York Times}, Aug 2025.
\newblock URL
  \url{https://www.nytimes.com/2025/08/18/world/europe/france-heat-wave-paris-climate-change-planning.html}.

\bibitem[Price et~al.(2024)Price, Sanchez{-}Gonzalez, Alet, et~al.]{Price2024}
I.~Price, A.~Sanchez{-}Gonzalez, F.~Alet, et~al.
\newblock Probabilistic weather forecasting with machine learning.
\newblock \emph{Nature}, 624:\penalty0 559--563, 2024.
\newblock \doi{10.1038/s41586-024-08252-9}.
\newblock URL \url{https://doi.org/10.1038/s41586-024-08252-9}.
\newblock Accessed 18 August 2025.

\bibitem[Ridgeway(2024)]{Ridgeway2024}
G.~Ridgeway.
\newblock Generalized boosted models: A guide to the gbm package.
\newblock 2024.
\newblock URL
  \url{https://cran.r-project.org/web/packages/gbm/vignettes/gbm.pdf}.

\bibitem[Romano et~al.(2019)Romano, Patterson, and Cand\`{e}s]{Romano2019}
Y.~Romano, E.~Patterson, and E.J. Cand\`{e}s.
\newblock Conformalized quantile regression.
\newblock In H.~Wallach et~al., editors, \emph{Advances in Neural Information
  Processing Systems 32 (NeurIPS 2019)}, volume~32, 2019.
\newblock URL
  \url{https://papers.neurips.cc/paper/8613-conformalized-quantile-regression.pdf}.

\bibitem[Rothfusz(1990)]{Rothfusz1990}
L.P. Rothfusz.
\newblock The heat index equation (or, more than you ever wanted to know about
  heat index).
\newblock Technical Report SR 90{-}23, National Weather Service, Southern
  Region Headquarters, Fort Worth, TX, 1990.
\newblock URL \url{https://www.weather.gov/media/ffc/ta_htindx.PDF}.
\newblock Scientific Services Division Technical Attachment.

\bibitem[Sarkar and Kuchibhotla(2023)]{Sarkar2023}
S.~Sarkar and A.K. Kuchibhotla.
\newblock Post-selection inference for conformal prediction: Trading off
  coverage for precision, 2023.
\newblock URL \url{https://doi.org/10.48550/arXiv.2304.06158}.

\bibitem[Schneider(1989)]{Schneider1989}
S.H. Schneider.
\newblock The greenhouse effect: Science and policy.
\newblock \emph{Science}, 243\penalty0 (4892):\penalty0 771--781, 1989.
\newblock \doi{10.1126/science.243.4892.771}.
\newblock URL \url{https://doi.org/10.1126/science.243.4892.771}.

\bibitem[Smith et~al.(2013)Smith, Zaitchik, and Gohlke]{Smith2013}
T.T. Smith, B.F. Zaitchik, and J.M. Gohlke.
\newblock Heat waves in the united states: Definitions, patterns and trends.
\newblock \emph{Climatic Change}, 118:\penalty0 811--825, 2013.
\newblock \doi{10.1007/s10584-012-0659-2}.
\newblock URL \url{https://doi.org/10.1007/s10584-012-0659-2}.

\bibitem[Steadman(1979)]{Steadman1979}
R.G. Steadman.
\newblock The assessment of sultriness. part i: A temperature-humidity index
  based on human physiology and clothing science.
\newblock \emph{Journal of Applied Meteorology}, 18\penalty0 (7):\penalty0
  861--873, 1979.
\newblock \doi{10.1175/1520-0450(1979)018<0861:TAOSPI>2.0.CO;2}.
\newblock URL
  \url{https://doi.org/10.1175/1520-0450(1979)018<0861:TAOSPI>2.0.CO;2}.

\bibitem[Stillman(2019)]{Stillman2019}
J.~H. Stillman.
\newblock Heat waves, the new normal: Summertime temperature extremes will
  impact animals, ecosystems, and human communities.
\newblock \emph{Physiology}, 34\penalty0 (2):\penalty0 861--873, 2019.
\newblock \doi{10.1152/physiol.00040.2018}.
\newblock URL \url{https://doi.org/10.1152/physiol.00040.2018}.

\bibitem[Stull(2017)]{Stull2017}
R.~Stull.
\newblock \emph{Practical Meteorology: An Algebra-Based Survey of Atmospheric
  Science}.
\newblock University of British Columbia Press, 2017.
\newblock URL \url{https://www.eoas.ubc.ca/books/Practical_Meteorology/}.

\bibitem[Tziperman(2022)]{Tziperman2022}
E.~Tziperman.
\newblock \emph{Global Warming Science}.
\newblock Princeton University Press, 2022.

\bibitem[Velthoen et~al.(2023)Velthoen, Dombry, Cai, and Engelke]{Velthoen2023}
J.~Velthoen, C.~Dombry, J.J. Cai, and S.~Engelke.
\newblock Gradient boosting for extreme quantile regression.
\newblock \emph{Extremes}, 26:\penalty0 639--667, 2023.
\newblock \doi{10.1007/s10687-023-00473-x}.
\newblock URL \url{https://doi.org/10.1007/s10687-023-00473-x}.

\bibitem[Walther et~al.(2002)Walther, Post, Convey, et~al.]{Walther2002}
G.-R. Walther, E.~Post, P.~Convey, et~al.
\newblock Ecological responses to recent climate change.
\newblock \emph{Nature}, 416\penalty0 (689):\penalty0 389--395, 2002.
\newblock \doi{10.1038/416389a}.
\newblock URL \url{https://doi.org/10.1038/416389a}.

\bibitem[Wang et~al.(2021)Wang, Tian, Lowe, Katlin, and Lehrter]{Wang2021}
F.~Wang, D.~Tian, L.~Lowe, L.~Katlin, and J.~Lehrter.
\newblock Deep learning for daily precipitation and temperature downscaling.
\newblock \emph{Water Resources Research}, 57\penalty0 (8):\penalty0
  e2020WR029308, 2021.
\newblock \doi{10.1029/2020WR029308}.
\newblock URL \url{https://doi.org/10.1029/2020WR029308}.

\bibitem[Warner(2004)]{Warner2004}
T.T. Warner.
\newblock \emph{Desert Meteorology}.
\newblock Cambridge University Press, Cambridge, 2004.
\newblock \doi{10.1017/CBO9780511535970}.
\newblock URL \url{https://doi.org/10.1017/CBO9780511535970}.

\bibitem[Xu et~al.(2014)Xu, Sheffield, Su, and et~al.]{Xu2014}
Z.~Xu, P.E. Sheffield, H.~Su, and et~al.
\newblock The impact of heat waves on children’s health: A systematic review.
\newblock \emph{International Journal of Biometeorology}, 58:\penalty0
  239--247, 2014.
\newblock \doi{10.1007/s00484-013-0655-x}.
\newblock URL \url{https://doi.org/10.1007/s00484-013-0655-x}.

\end{thebibliography}

\end{document}